\documentclass[prb,twocolumn,showpacs]{revtex4}
\usepackage{graphicx}
\usepackage{dcolumn}
\usepackage{bm}

\begin{document}

\preprint{APS/123-QED}

\title{Paramagnetic and antiferromagnetic resonances\\ in the diamagnetically
diluted Haldane magnet PbNi$_2$V$_2$O$_8$}

\author{A.~I.~Smirnov}
\author{V.~N.~Glazkov}
\affiliation{P.~L.~Kapitza  Institute for  Physical  Problems RAS,
117334 Moscow, Russia}

\author{H.-A.~Krug von Nidda }
\author{A.~Loidl}
\affiliation{Experimentalphysik V, Electronic Correlations and
Magnetism, Institute of Physics,\\ University of  Augsburg,
D-86135 Augsburg, Germany}
\author{L.~N.~Demianets}
\author{ A.~Ya.~Shapiro}
\affiliation{A.~V.~Shubnikov  Institute for  Crystallography  RAS,
117333 Moscow, Russia}
\date{\today}

\begin{abstract}
The impurity-induced antiferromagnetic ordering of the doped
Haldane magnet Pb(Ni$_{1-x}$Mg$_{x}$)$_2$V$_2$O$_8$ $(0\leq x \leq
0.06)$ was studied by electron spin resonance (ESR) on ceramic
samples in the frequency range 9-110~GHz. Below the N\'{e}el
temperature a transformation of the ESR spectrum was found,
indicating an antiferromagnetic resonance mode of spin precession.
The excitation gap of the spin-wave spectrum increases with
increasing Mg-concentration $x$ in the same manner as the N\'{e}el
temperature, reaching its maximum value of 80 GHz at
$x$~$\geq$~0.04. At small concentrations $x <$ 0.02 the signals of
antiferromagnetic resonance were found to coexist with the signal
of the paramagnetic resonance indicating a microscopic separation
of the magnetic phases.
\end{abstract}


\maketitle

\section{Introduction}

One dimensional magnetic systems exhibit a variety of different
collective quantum states which are of significant physical
interest. The ground state and magnetic properties of integer and
half-integer 1D-spin systems are quite different. The ground state
of the spin $S=1$ chain with  antiferromagnetic exchange is a
singlet, and the excited triplet states are separated by an energy
gap $\Delta$ (so called Haldane gap).\cite{Haldane} The uniform
$S=1/2$ antiferromagnetic spin chain exhibits also a singlet
ground state but a gapless spectrum of the triplet excitations.
\cite{desCloizeaux} Both $S=1$ and $S=1/2$ spin chains are
disordered in the ground state: the average spin projection
$\langle S_i^z \rangle$ equals zero for all spins of the chain.  Nevertheless,
there is  a difference in the magnetic correlation length $\xi$
which is infinite for the uniform $S=1/2$ chain and finite for the
Haldane chains.

 A spin-gap opens also in the spectrum of $S=1/2$ chains, if the chains
are alternated, i.e.  the intrachain-exchange integral takes in
turn two values $J \pm \delta J$.  The correlation length of an
alternated chain  becomes finite.  The alternating of spin chains
may occur due to the crystal structure as, e.g., in the 1D magnet
(VO)$_2$P$_2$O$_7$, \cite{VOPO} or due to the spin-Peierls
transition \cite{Pytte} as, e.g., in CuGeO$_3$. \cite{Hase} The
magnetic behavior of spin-gap systems is marked by a
characteristic freezing out of the magnetic susceptibility at low
temperatures $T < \Delta/k_B$.

The impurities embedded in a spin-gap magnet provide magnetic
degrees of freedom on the singlet  background and  result in a
contribution to the magnetic susceptibility at low temperatures.
The analysis of spin-Peierls \cite{Fukuyama,Khomskii} and of
Haldane \cite{Miyashita} systems shows that breaking of spin
chains locally destroys the singlet state, and multi-spin clusters
with a local staggered magnetization (nonzero
$\langle S_i^z \rangle$) occur
in the vicinity of impurity atoms, on both sides of the
diamagnetic impurity.  The staggered magnetization decays
exponentially with the distance apart from the impurity atom, thus
the cluster includes spins placed on the chain part of the length
of about of $\xi$ on one side of the impurity.  These clusters
have a net spin and a net magnetic moment.  For a broken dimerized
$S=1/2$ chain the net spin value  of a cluster is obviously 1/2.
An intriguing suggestion was made for spin $S=1$ chains stating
that spin vacancies should result in effective $S=1/2$ degrees of
freedom at the ends of the long chain segments \cite{s_one_half}.
ESR experiments confirmed the multi-spin nature of the magnetic
defects created by impurities in the spin-Peierls magnet
CuGeO$_3$. \cite{Glazkov,Glazkov2} The indications of $S=1/2$
degrees of freedom were  found in ESR spectra of the doped Haldane
system NENP \cite{s_one_half,Clarum} and for  defects in the
Haldane magnet Y$_2$BaNiO$_5$, \cite{Kimura} while the $S=1/2$
effective spins were questioned in Ref.~\onlinecite{Ramirez} on
the base of specific-heat measurements.  A discussion of the
$S=1/2$ problem may be found in Ref.~\onlinecite{Batista,Batista2}
and in the review article Ref.~\onlinecite{Katsumata}.

Another interesting effect associated with defects in spin-gap
systems is impurity-induced magnetic ordering.  Doping of the
spin-Peierls compound CuGeO$_3$ with both magnetic and nonmagnetic
impurities and the resulting antiferromagnetic ordering was
extensively studied (see, e.g.,
Refs.~\onlinecite{Regnault,Renard,Grenier,Masuda,Glazkov3}), while
a spin-vacancy induced magnetic ordering in a Haldane magnet was
only recently reported  \cite{Uchiyama} for PbNi$_2$V$_2$O$_8$.
The phenomenon of impurity-induced antiferromagnetic ordering  in
a spin-Peierls magnet was explained by the overlap of the wings of
the soliton-like clusters of the staggered magnetization and the
interchain exchange.  \cite{Fukuyama,Khomskii} For Haldane systems
an impurity-induced ordering was predicted taking into account the
restoring of antiferromagnetic correlations by impurities.
\cite{ShenderKivelson} Both approaches have the same physical
reason: the staggered magnetization is restored over a number of
spins near the defect.  The evolution from the isolated spin
clusters to the long-range ordered system was followed by
observations of ESR signals for the doped spin-Peierls magnet
CuGeO$_3$. \cite{Glazkov,Glazkov2,Glazkov3} The gap in the
spin-wave spectrum and other characteristics of the
impurity-stimulated antiferromagnetic phase were derived from the
spectrum of the antiferromagnetic resonance (AFMR).
\cite{Glazkov,Hase2,Smirnov,Nojiri}

The goal of the present work is an ESR study of defects and a
search for the AFMR signals in the doped Haldane magnet
PbNi$_2$V$_2$O$_8$. The diamagnetic dilution is performed
following Ref.~\onlinecite{Uchiyama} via the substitution of a
small part of $S=1$ Ni$^{2+}$ ions by $S=0$ Mg$^{2+}$ ions.   As a
result we observed ESR signals from magnetic defects created by
spin vacancies.  Above the N\'{e}el point the ESR spectrum
corresponds to the respective $S=1/2$ degrees of freedom. At the
N\'{e}el point the ESR spectrum observed for high concentration of
impurities exhibits a transformation to an antiferromagnetic
resonance spectrum of a uniaxial antiferromagnet. At low
concentrations ($x\leq 0.02$) we observe the coexistence of
paramagnetic  and antiferromagnetic resonance. We believe that
this coexistence reveals the microscopic separation of
antiferromagnetic and paramagnetic phases  as in the case of the
spin-Peierls system CuGeO$_3$.\cite{Glazkov3}

\section{Samples and experimental details}

Solid solutions of Pb(Ni$_{1-x}$Mg$_{x}$)$_2$V$_2$O$_8$  with
$x=0,~0.01,~0.02,~0.04,~0.06$ were prepared by means of ceramic
technology using oxides PbO, NiO, MgO and V$_2$O$_5$ of high
purity as starting chemical reagents.  Annealing was performed at
1050~K in air. At higher temperatures an incongruent melting of
PbNi$_2$V$_2$O$_8$ occurs. Single-phase samples were produced by
treating the initial mixture at high temperatures for 800 hours
with several intermediate grindings and pressings. The sample
composition was controlled by x-ray diffraction using  the Dmax
IIIC diffractometer of Rigaku. According to the x-ray diffraction
data, the  single-phase samples have tetragonal structure of space
group I4$_1$cd and are isostructural with SrNi$_2$V$_2$O$_8$. The
cell parameters of the pure compound are in accordance with those
observed in Ref.~\onlinecite{Uchiyama}.  Partial substitution of Ni
by Mg results in a compression of the elementary cell with
increasing Mg concentration.  The change of the unit-cell
parameters with doping is presented in the Table
~\ref{tab:table1}.

\begin{table}
\caption{\label{tab:table1}  Lattice parameters and unit-cell
volume of Pb(Ni$_{1-x}$Mg$_{x}$)$_2$V$_2$O$_8$ for Mg
concentrations $0 \leq x \leq 0.06$
 }
\begin{ruledtabular}
\begin{tabular}{cccc}
$~x~$ &$~a~$, \AA~~& $c$, \AA~~~&$V$, \AA$^3$
\\ \hline
0&12.251(3)&8.353(4)&1253.6(4) \\
0.01&12.244(2)&8.349(3)&1251.8(3)\\
0.02&12.230(3)&8.341(3)&1247.7(4) \\
0.04&12.228(3)&8.321(3)&1241.9(8) \\
0.06&12.205(3)&8.315(3)&1238.5(4) \\
\end{tabular}
\end{ruledtabular}
\end{table}

The molar concentration of impurity phases is estimated to be less
than 1\%.

The temperature dependences of the susceptibility were measured
with a SQUID magnetometer (Quantum Design) in the range
1.8~--~400~K. The ESR spectra were measured in a spectrometer with
a set of transmission-type  resonators covering the range from 18
to 110 GHz and for temperatures 1.3~--~30~K.  Magnetic resonance
absorption lines were recorded as a function of the transmitted
microwave power on the applied magnetic field.  The reduction   of
the transmitted signal is proportional to the microwave power
absorbed by the sample.  In addition X-band (9.5~GHz) measurements
were performed with a Bruker ELEXSYS E500 CW spectrometer.

\section{Experimental results}

\begin{figure}
\includegraphics[width=\columnwidth]{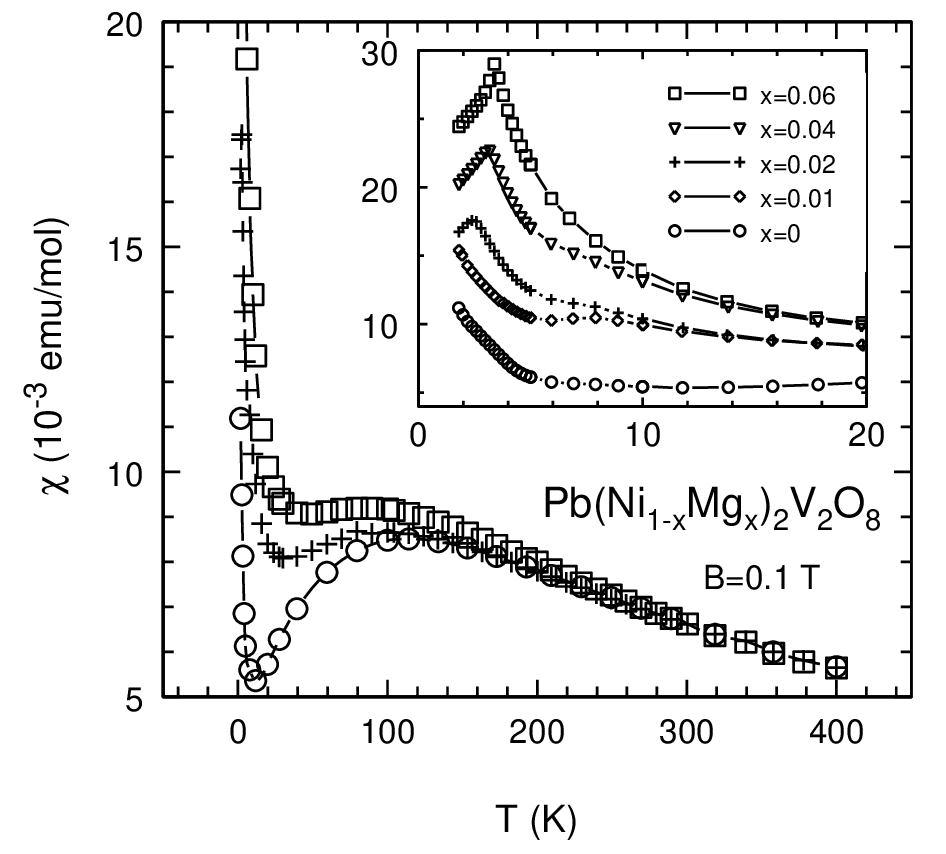}
\caption{\label{fig:f1}Temperature dependences of the
susceptibility  of ceramic Pb(Ni$_{1-x}$Mg$_{x}$)$_2$V$_2$O$_8$
for $0 \leq x \leq 0.06$. Inset: susceptibility below 20~K.}
\end{figure}

The temperature dependence of the susceptibility for several
samples is shown in the Fig.~\ref{fig:f1}. The susceptibility of
the  pure ($x=0$) compound  exhibits a broad maximum at $T=120$~K
and decreases with decreasing temperature. At $T=10$~K the
susceptibility reveals a minimum and than shows a Curie-like
increase due to defects. Below $T=$~10~K the total susceptibility
of the nominally pure sample can be accounted for one percent of
free Ni-ions.

Doping with Mg results in an increase of the susceptibility which
is proportional to the concentration of impurities $x$.  For
one-percent doping  the susceptibility at $T<10$~K amounts about
two times the  susceptibility of the pure sample.  For the samples
with the concentration  $x=$~0.02, 0.04 and 0.06  we additionally
observed kinks in the susceptibility curves indicating the
N\'{e}el transition temperature $T_N$ in a good accordance with
Ref.\onlinecite{Uchiyama}.  Some of the samples demonstrate a
nonmonotoneous contribution to the susceptibility with a week
smeared-out anomaly close to 8~K. A corresponding anomaly is
observed in the temperature dependence of the microwave absorption
(see below at the end of the Section), but is suppressed in
magnetic fields larger than 0.2~T.  The magnitude of this anomaly
depends on the number of intermediate grindings and annealings and
probably is due to onset of superconductivity within a parasitic
plumbous phase.

\begin{figure}
\includegraphics[width=\columnwidth]{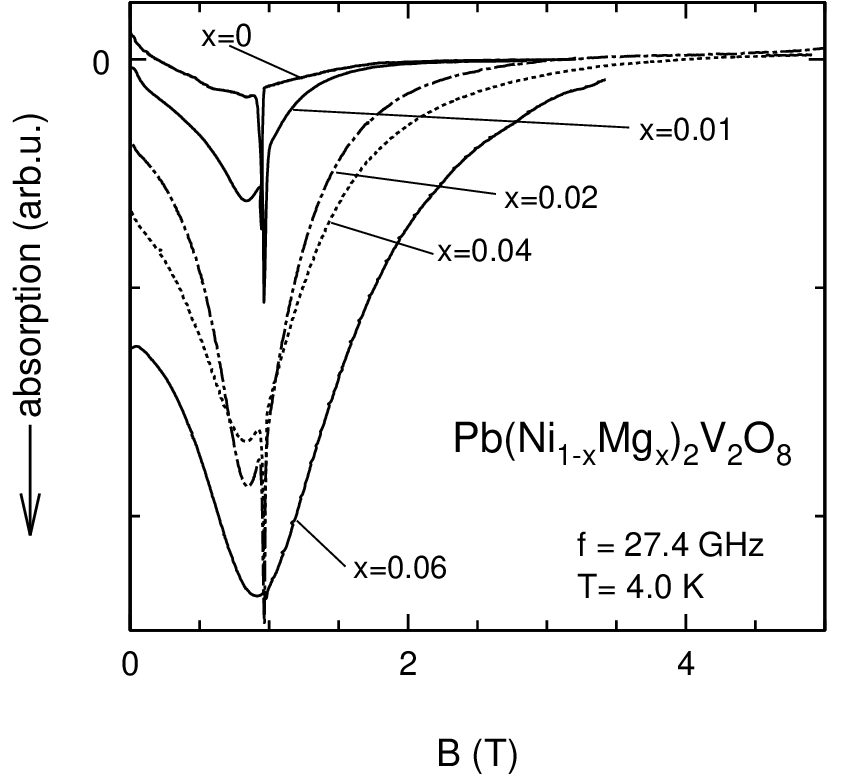}
\caption{\label{fig:f2}ESR  spectra of ceramic
Pb(Ni$_{1-x}$Mg$_{x}$)$_2$V$_2$O$_8$ for $0 \leq x \leq 0.06$ at
the frequency 27.4~GHz  at $T=$~4~K.  Narrow lines are DPPH-labels
corresponding to $g=$~2.0.}
\end{figure}

The ESR lines at $T=4$~K (this temperature is above T$_N$ for all
samples) are plotted in Fig.~\ref{fig:f2}. The relatively weak ESR
signal in the nominally pure sample exhibits a complicated
structure consisting of more than four spectral components.  In
doped samples  in the paramagnetic state the ESR absorption is
much more intensive and its spectrum consists of a single line at
the position corresponding to a $g$-factor of 2.2 in the frequency
range 9~--~110~GHz.

\begin{figure}
\includegraphics[width=\columnwidth]{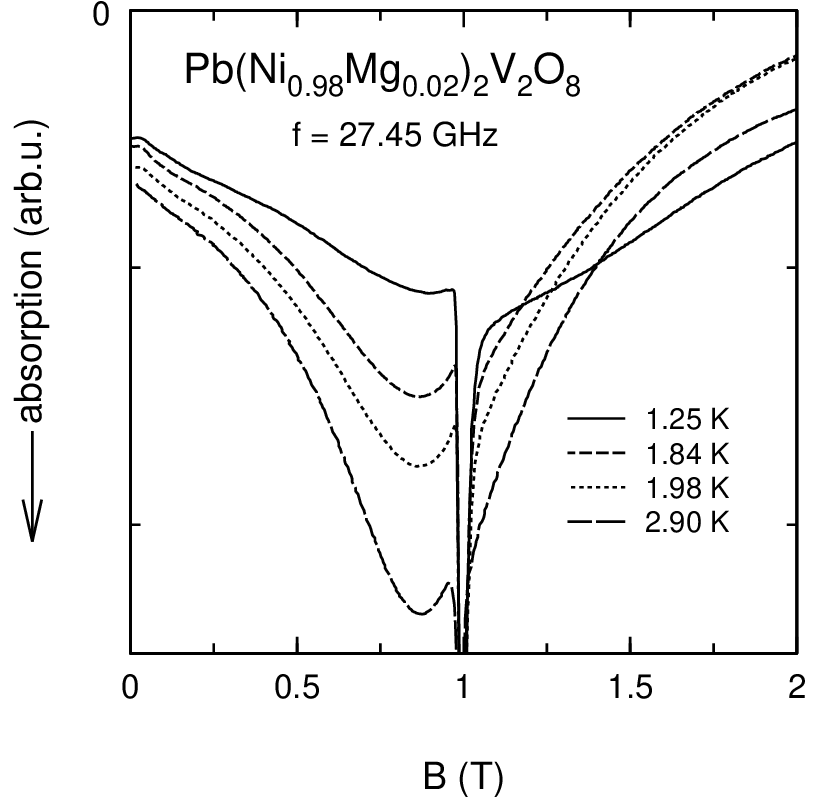}
\caption{\label{fig:f3} Temperature evolution of the 27.3~GHz ESR
lines for $x=$~0.02.}
\end{figure}

\begin{figure}
\includegraphics[width=\columnwidth]{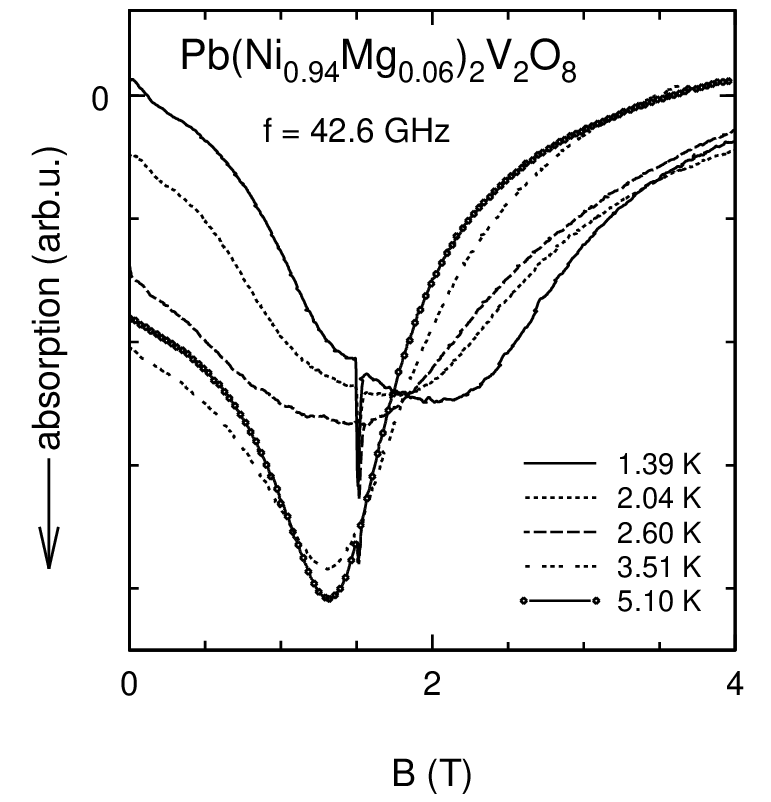}
\caption{\label{fig:f4} Temperature evolution of the 42.6~GHz ESR
lines for $x=$~0.06.}
\end{figure}

\begin{figure}
\includegraphics[width=\columnwidth]{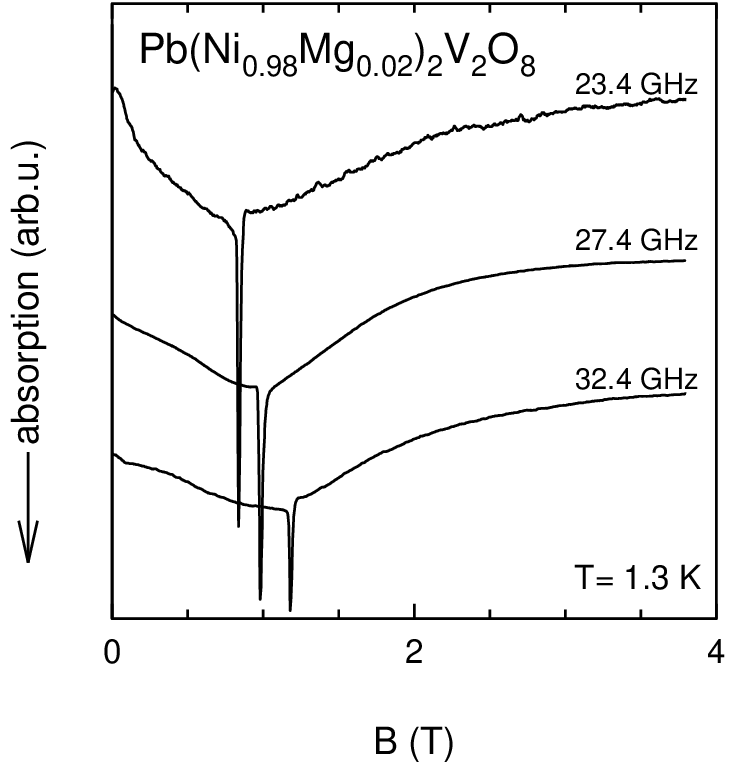}
\caption{\label{fig:f5} ESR lines of the sample with $x=$~0.02 for
$T=1.3$~K at different frequencies. The curves are shifted
vertically.}
\end{figure}

\begin{figure}
\includegraphics[width=85mm]{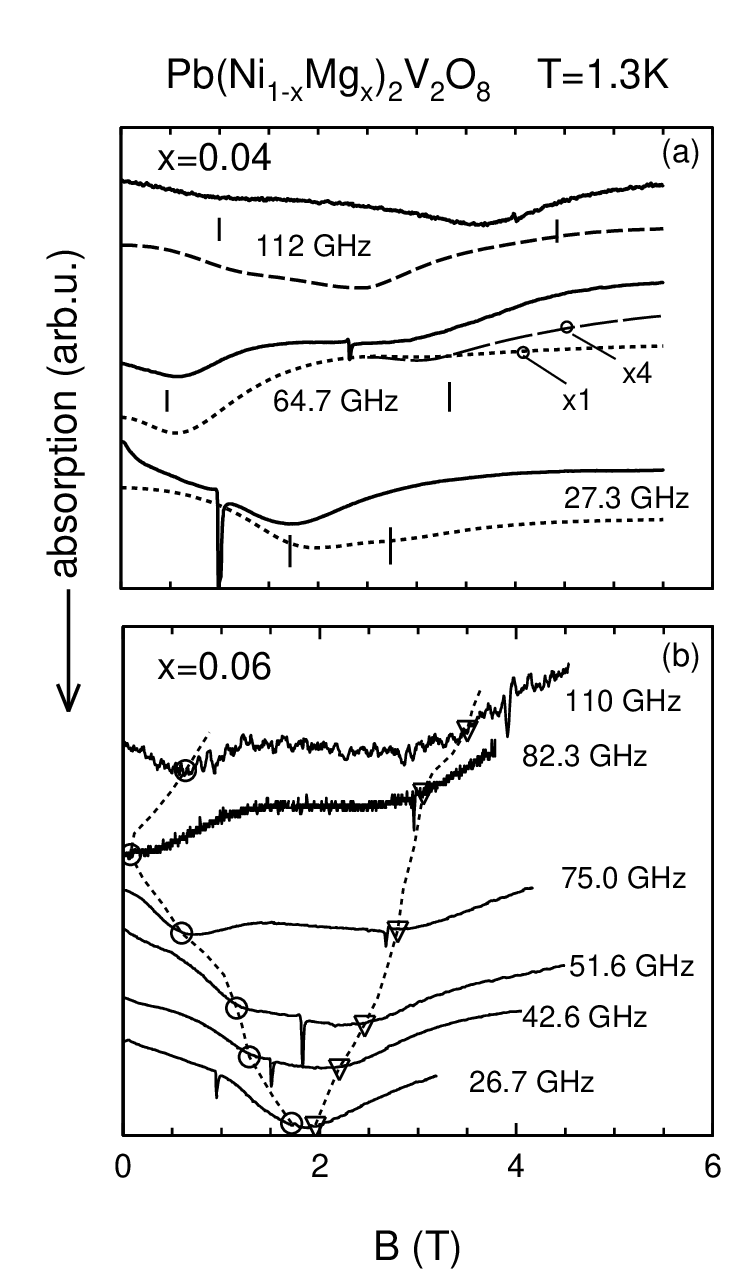}
\caption{\label{fig:f6} ESR lines of the samples with $x=$~0.04
(panel (a)) and  $x=$~0.06 (panel (b)) for $T=1.3$~K at different
frequencies. The dashed curves on panel (a) present the
calculated absorption of an antiferromagnetic  powder with the
following parameters: $\omega_0/2\pi=$~80~GHz, linewidths of the
upper and lower AFMR branches were chosen equal to
0.2$\omega_0/2\pi$, $\gamma /2\pi= $~3.1~GHz/kOe corresponds to
the $g$-factor 2.2. Vertical segments mark the calculated
boundaries of absorption at zero AFMR linewidth.  The circles and
triangles on the curves of the panel (b) demonstrate the
 lower and upper boundaries of the absorption bands, respectively. The
 dashed lines connecting these signs are guide to eyes.}
\end{figure}

The temperature evolution of the ESR spectra of samples with
different doping concentrations is shown in Figs.~\ref{fig:f3} and
\ref{fig:f4}.  One can see that below T$_N(x)$ ($T_N$ is 2.4~K at
$x=$~0.02 and $3.4$~K at $x=$~0.06) a transformation of the
spectral density occurs: the maximum of the microwave absorption
shifts to higher fields with respect to the resonance field of the
paramagnetic phase. The ESR line transforms into a broad band of
absorption. The transition temperature for $x = 0.01$ is on the
edge of the temperature range of the susceptibility measurements
and was determined as 1.8~K from the appearance of ESR-line
distortion and from the microwave anomaly in zero field described
below. The transformation of the ESR line at $T<T_N$ is
accompanied by a strong reduction of the absorption at the field
of the paramagnetic resonance for $x = 0.04$ and $0.06$, while for
$x = 0.01$ and $0.02$ the absorption in this field remains
comparable with that in the paramagnetic phase. With increasing
frequency the band of absorption becomes wider as documented in
Figs.~\ref{fig:f5} and \ref{fig:f6}. Although at $x = 0.01$ and
$0.02$ the local maximum of the ESR absorption remains at the
paramagnetic position (see Fig.~\ref{fig:f3}), nevertheless, at
$T<T_N$ the line-shape changes from a Lorentzian to a line with
additional absorption in the wings. As documented in
Fig.~\ref{fig:f5}, there is an enlarged absorption on the right
wing of the ESR line at the lower frequencies of 23.4 GHz and 27.4
GHz and an enlarged absorption on both wings for the higher
frequency of 32.4 GHz.

\begin{figure}
\includegraphics[width=\columnwidth]{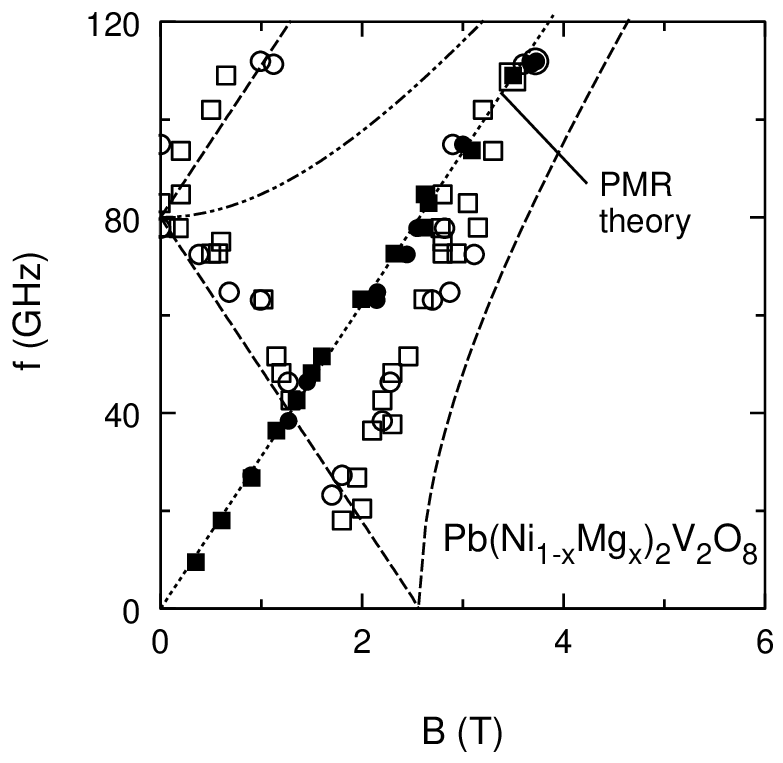}
\caption{\label{fig:f7} Frequency-field diagram of the microwave
absorption in Pb(Ni$_{1-x}$Mg$_{x}$)$_2$V$_2$O$_8$. Open symbols
are boundaries of the AFMR absorption bands at $T=1.3$~K:  squares
- $x=0.06$, circles - $x=0.04$.  Full symbols are resonance fields
in the paramagnetic state, taken at 4~K:   squares - $x=0.06$,
circles - $x=0.04$. Dashed lines are theoretical dependencies for
AFMR frequencies given by formulae 1-3, the dotted line is the
paramagnetic resonance dependence for spins $S=1/2$ with
$g=$~2.22}
\end{figure}

For the 0.04 and 0.06 samples
the boundaries of the absorption band become clearly visible (see
Figs.~\ref{fig:f6}, panels (a) and (b) respectively).  They were
determined from the field values at the left and right edges of
the band, where the absorption stops to grow or begins to
diminish, respectively. These boundaries are marked in
Fig.~\ref{fig:f6}~(b) by circles and triangles connected with
dashed lines. The frequency-field dependences of the left and
right boundaries of the absorption are plotted in
Fig.~\ref{fig:f7} along with the frequency-field dependence of the
ESR in the paramagnetic phase at 4~K.  The absorption-band edges
for $x = 0.04$ and $0.06$ exhibit a pattern typical for a powder
of an antiferromagnet (see, {\it e.g.} Ref.~\onlinecite{Smirnov2})
revealing an antiferromagnetic resonance gap of 80~GHz.

\begin{figure}
\includegraphics[width=\columnwidth]{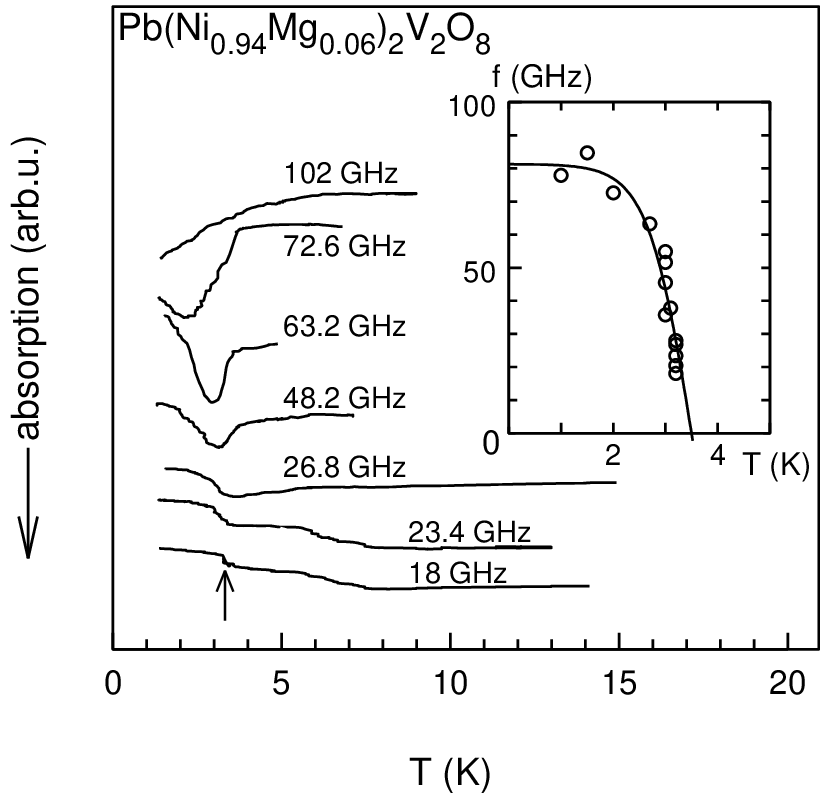}
\caption{\label{fig:f8} The dependences of the microwave power
transmitted through the resonator on the temperature at $B=$~0 for
the sample doped with 0.06~Mg.  Inset: temperature dependence of
the antiferromagnetic resonance gap at $x=$~0.06.}
\end{figure}

\begin{figure}
\includegraphics[width=\columnwidth]{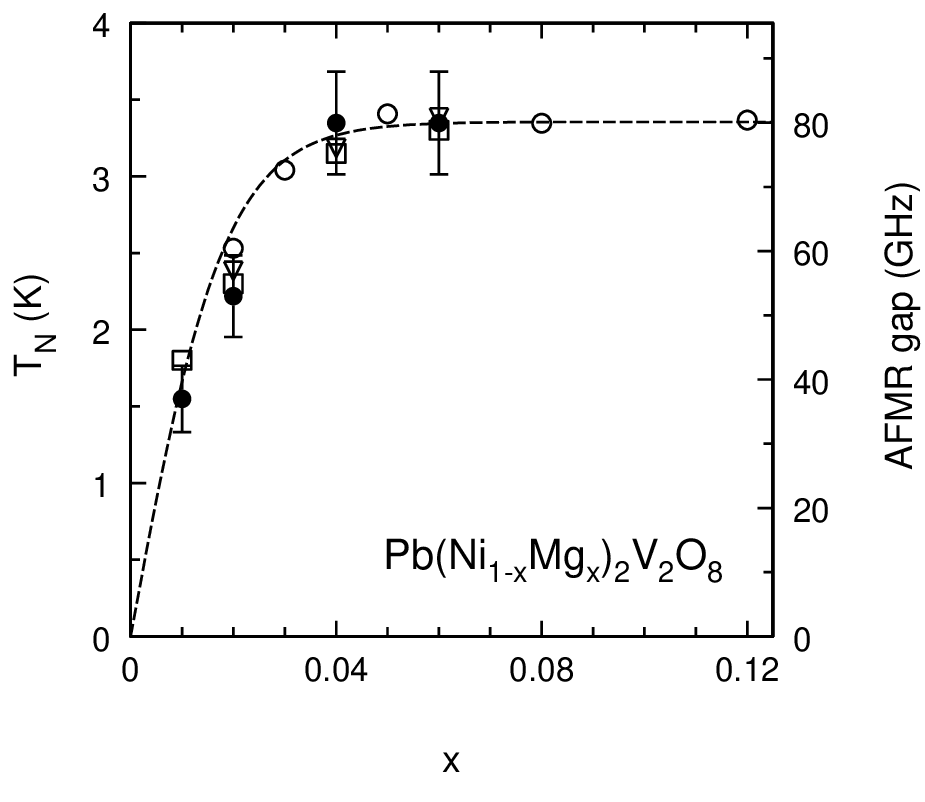}
\caption{\label{fig:f9} The concentration dependence of the
N\'{e}el temperature and of the antiferromagnetic resonance gap.
The meaning of the symbols is as follows: squares - $T_N$ from
microwave absorption, open circles -$T_N$ from
Ref.~\onlinecite{Uchiyama}, triangles -$T_N$ from {\it
dc}-susceptibility, full circles - AFMR gap. The lines are drawn
to guide the eye.}
\end{figure}

 The AFMR gap  should be
 temperature dependent being zero at $T=T_N$ and reaching its maximum
 value at $T=0$.  This fact enables the observation of the
antiferromagnetic resonance in zero field by recording the
microwave absorption {\it vs.} temperature.  The temperature
dependences of the transmitted microwave power at different
frequencies are shown in Fig.~\ref{fig:f8} for $x = 0.06$.  The
inset of Fig.~\ref{fig:f8} shows the dependence of the zero-field
resonance frequency on temperature determined from the maximum
absorption at each frequency. The extrapolation of this frequency
to zero temperature allows one to obtain the AFMR gap
$\omega_0/2\pi$ at $T=0$. The abrupt change, which is visible in
the temperature dependence of the microwave absorption at low
frequencies $f \ll \omega_0/2\pi$, marks  the N\'{e}el temperature
(see the arrow on Fig.~\ref{fig:f8}). The values of $T_N$ obtained
by this method are  in a good agreement with the results of {\it
dc}-susceptibility for all concentrations.
The values of the N\'{e}el temperatures and antiferromagnetic
resonance gaps for samples with different $x$ are shown in
Fig.~\ref{fig:f9}. Both values increase approximately linear for
$x < 0.04$ and saturate above $x=$~0.04

\section{Discussion}

\subsection{Residual and substitutional defects}

In our investigation we observed that the substitution of Ni- with
Mg-ions strongly changes  the magnetic susceptibility and the ESR
spectra of PbNi$_2$V$_2$O$_8$.  The influence of inevitable
impurity phases can be estimated from the residual low-temperature
susceptibility which should tend to zero in a perfect Haldane
magnet. The low-temperature susceptibility provided by these
phases equals in magnitude the susceptibility due to one-percent
Mg-doping and is definitely smaller than the susceptibility of the
higher doped samples.

The integrated ESR intensity is a measure of the spin
susceptibility in the paramagnetic state.  The ESR signal of
residual impurities observed in the nominally pure sample at low
temperatures is two times smaller than the intensity which results
from doping with 1\% Mg. Thus the spin susceptibility  provided by
Mg doping prevails over the spin susceptibility of impurity
phases.

 The weak ESR signal of the residual defects has a complicated
spectrum with four lines which are resolved at $T=$~1.4~K  and
which are smeared into one very broad line at $T> $~25~K.
Analogous to Ref.~\onlinecite{Kimura} we ascribe this signal
presumably to the Ni$^{3+}$-ions appearing due to oxygen
nonstoichiometry.

\subsection{Paramagnetic resonance}

At 4~K the paramagnetic resonance frequency of all doped samples
is proportional to the magnetic field for frequencies 9 GHz $\leq
f \leq$ 110 GHz, as shown on Fig.~\ref{fig:f7} for concentrations
$x = 0.04$ and $0.06$. We do not observe a zero-field ESR
splitting which is usually expected for spin $S=1$ in a crystal
field and which is absent for spin $S=1/2$. From the data shown on
Fig.~\ref{fig:f7} we derive an upper limit for the zero-field
splitting of 0.5~GHz. The absence of a zero-field ESR splitting
strongly confirms that the Mg-doping provides effective $S=$1/2
degrees of freedom.  The expected value of the zero-field
splitting for the single Ni$^{2+}$-ion is determined by the
crystal-field parameter $D$ of the spin-Hamiltonian single-ion
anisotropy term $D(S_{i}^z)^2$. For PbNi$_2$V$_2$O$_8$ the value
of $D$ was derived from the spectrum of spin excitations:
$D=-0.23$~meV~ \cite{Uchiyama} and $D=-0.45$~meV. \cite{Zheludev}
The discrepancy between these two values is due to different
models of the magnetic interactions used for the data analysis.
Even the smaller value should result in a zero-field splitting of
the spin $S=1$ resonance mode of about 50~GHz. For the case of the
ceramic sample it would result in a wide band of absorption which
is not compatible with the observed resonance lines shown in
Fig.~\ref{fig:f2}. The effect of the single-ion anisotropy on the
energy spectrum of the end of the Haldane chain was analyzed in
Ref.~\onlinecite{Batista2}, were it was shown that the $D$-term
does not affect the central peak frequency of the ESR spectrum of
long spin chain segments, and thus the spectrum is consistent with
the concept of the effective spin $S=1/2$. Our observations
confirm this conclusion.

The samples of PbNi$_2$V$_2$O$_8$ allow a relatively high extent
of doping in comparison with other Haldane magnets (see, e.g.
Refs.~\onlinecite{s_one_half,Batista2}). This relatively high
concentration of dopants enables  to detect the interaction
between defects by  a comparison of the linewidth at different
$x$. From the ESR lines displayed on Fig.~\ref{fig:f2} we see that
the linewidth at $x=$~0.02 is about 0.5~T larger than at
$x=$~0.01. This huge increase cannot be ascribed to dipole-dipole
interaction or to exchange interaction of neighboring spins at
these concentration values. Taking into account the multi-spin
nature of the magnetic defects we can explain this increase as
follows. The staggered magnetization exists on the chain areas
with the length of about $\xi$ at the ends of the chain segments.
At small concentrations the portion of spin-chain segments which
are shorter than $2\xi$ equals approximately $2 x \xi$ (here $\xi$
is measured in interspin distances). Thus at the reasonable value
of $\xi \approx $~10 a significant portion of chain segments
(about 0.4 at $x=$~0.02) should contain strongly interacting
clusters, touching each other.  This interaction may cause the
broadening of the ESR line because the exchange interaction along
with the single-ion anisotropy results in an effective anisotropic
exchange giving rise to the line broadening.

\subsection{Antiferromagnetic resonance}

By cooling through the N\'{e}el temperature we observed a
significant transformation of the ESR spectra. The conversion of a
single ESR line into a  band of absorption in ceramic samples may
be explained by the opening of an energy gap in the
antiferromagnetic resonance spectrum and by the dependence of the
AFMR frequency on the orientation of the magnetic field.  The
observation of the "temperature resonance" by scanning the
temperature confirms the opening of the AFMR gap at $T=T_N$.

The antiferromagnetic spin structure on the tetragonal lattice
should be of uniaxial anisotropy, therefore we fitted the
boundaries of the absorption bands (Fig.~\ref{fig:f7})  by the
formulae for AFMR frequencies of a uniaxial antiferromagnet:
\cite{AFMR}

{\it a)}
 $H \parallel$
easy axis, $H<H_c=\omega_0/\gamma$:

\begin{equation}
\omega_{1,2}= \omega_0 \pm \gamma H.
\end{equation}

{\it b)} $H \parallel$ easy axis,  $H>H_c$:
\begin{eqnarray}
\omega_{1}=0,\\ \nonumber \omega_2=\sqrt{(\gamma H)^2-
\omega_0^2}.
\end{eqnarray}

{\it c)} $H \perp$ easy axis:
\begin{eqnarray}
\omega_{1}= \omega_0,\\ \nonumber \omega_2=\sqrt{(\gamma H)^2+
\omega_0^2}.
\end{eqnarray}

Here $\gamma=2 \mu_B/\hbar$. For intermediate  orientations the
resonance frequencies lie between the values given by Eqns.~(1-3).
Thus, the magnetic fields determined by these formulae represent
the boundaries of the absorption bands of the powder sample at a
given frequency. The boundaries calculated in this way are shown
in Fig.~\ref{fig:f7} along with the boundaries derived from the
observed absorption for the samples with $x = 0.04$ and $0.06$.

The resonance frequencies for an arbitrary orientation of the
magnetic field can also be calculated.\cite{AFMR} We modeled the
field dependence of the absorption of a ceramic sample by
integration of the absorption over all orientations of the
crystallites, taking into account the finite ESR linewidth and
orientational dependence of AFMR frequencies  and neglecting the
dependence of the ESR susceptibility of both AFMR modes on the
orientation of the vector of antiferromagnetism with respect to
the magnetic microwave field. This neglecting may result in a loss
of a nonresonant but field-dependent factor.  The results of this
calculation are plotted on Fig.~\ref{fig:f6}(a) along with the
experimental curves. The modeling at finite linewidths shows the
smeared boundaries of the absorption bands near the field values
calculated by Eqns.~(1-3) which indicate boundaries of bands in
the limit of narrow  AFMR modes.  These zero-width boundaries are
marked on Fig.~\ref{fig:f6}(a) by vertical segments.

We find qualitative agreement between the experimental results and
the calculated AFMR frequencies plotted on Fig.~\ref{fig:f7} and
between experimental and modeled absorption curves on
Fig.~\ref{fig:f6}. The comparison of modeled and experimental
spectra allows us to estimate the AFMR linewidth as approximately
20~GHz. Due to this large linewidth the observed AFMR absorption
deviates from the absorption of the usual antiferromagnetic
powder, \cite{Smirnov2} for which the boundaries of absorption
were observed as sharp anomalies at the limiting positions
calculated from formulae like given above.

\subsection{Coexistence of two ESR modes in low-doped samples}

The ESR spectra  of the low-doped samples with $x = 0.01$ and $x =
0.02$ do not allow to obtain the frequency-field diagram of the
boundaries of the absorption band, because the paramagnetic
resonance line dominates within the complete frequency range. The
absorption curve looks like a broad resonance line rather than an
absorption band. Nevertheless the gap values can be determined
from the absorption {\it vs.} temperature dependencies in the zero
field.  The observed absorption at the paramagnetic resonance
field is definitely different from the AFMR absorption, because at
the AFMR gap of 51~GHz the AFMR field for the 18~GHz measurement
should be 1.9~T and not at 0.6 T as observed. In addition the
field of the maximum of absorption is proportional to the
frequency (see Fig.~\ref{fig:f5}).  This proportionality
distinctly distinguishes the paramagnetic resonance mode from the
AFMR mode. The presence of a paramagnetic resonance line of
sufficient intensity (see Fig.~\ref{fig:f5}) is not due to an
inhomogeneity of the impurity concentration. An inhomogeneous
impurity concentration should smear the N\'eel point and result in
macroscopic antiferromagnetic and paramagnetic parts of the
sample.  Our susceptibility curves demonstrate a well defined
antiferromagnetic transition with a width less than 0.2~K, whereas
we observe a paramagnetic ESR line at $T_N-T
> 0.8$~K.

This coexistence of  paramagnetic and antiferromagnetic resonance
modes is related to the recent observation of a coexistence of
these modes in the impurity-stimulated antiferromagnetic phase of
the spin-Peierls magnet CuGeO$_3$.\cite{Glazkov3,Smirnov}
Unfortunately in the present work we cannot distinguish between
the AFMR and paramagnetic modes so clearly as in the  case of
single crystals of doped CuGeO$_3$ with narrow ESR lines. The
coexistence of these two kinds of resonance signals was explained
in Ref.~\onlinecite{Glazkov3} taking into account the random
distribution of the impurities and a respective dispersion of the
length of spin-chain segments. As a result, there are relatively
large antiferromagnetic areas, formed by coupled clusters, and
isolated spin clusters, separated from each other by the
nonmagnetic spin-gap phase. The isolated spin clusters (described
in the Introduction) provide a paramagnetic resonance signal due
to their net magnetic moment, while larger antiferromagnetic areas
give rise to AFMR signals. This scenario of the microscopic phase
separation may be derived also from the  numerically simulated 2D
pattern of the stimulated staggered magnetization at $T=0$
obtained both for spin-Peierls and Haldane magnets with an
interchain interaction in Ref.~\onlinecite{Yasuda}.  There are
peaks of staggered magnetization around impurities and areas of a
much lower order parameter between them. Assuming naturally that
at finite temperature the areas of low order parameter will be
disordered, we arrive at a pattern of islands of different sizes
with local antiferromagnetic order.

\subsection{The average spin and the
N\'eel temperature  of the impurity induced order}

The  concentration-dependences of the AFMR gap and the N\'eel
temperature are similar in the investigated range of
concentrations. They are approximately linear at small $x$ and
saturate at $x \approx 0.04$.   The linear increase of the AFMR
gap and of the N\'eel temperature at low $x$ may be explained
using the concept of "magnetic molecules" \cite{ShenderKivelson}
appearing near impurities. According to this consideration  we
take a cluster with the staggered magnetization appearing near an
impurity as a "magnetic molecule" with the number of spins of the
order of the correlation length $\xi$. The  module of the spin
projection averaged over the sample is estimated as follows:

\begin{equation}
\langle|S_{i}^z|\rangle \sim x \xi S . \label{averS}
\end{equation}

The linear $x$-dependence of $\langle|S_{i}^z|\rangle$ at the diamagnetic
dilution was also  confirmed by the Monte-Carlo simulation.
\cite{Yasuda}

Further, analyzing the interaction of the magnetic molecules via
the interchain exchange $J_\perp $ the following estimation for
the  Curie-Weiss-constant was found in the low concentration
limit: \cite{ShenderKivelson}

\begin{equation}
\theta \sim x J_\perp (\xi S)^2 \sim J_\perp \langle|S_{i}^z|\rangle \xi S
\label{theta}
\end{equation}

This estimate agrees well with the observed linear dependence of
the N\'eel temperature on the concentration. In the
molecular-field approximation the AFMR gap is determined by the
expression

\begin{equation}
\hbar \omega_0=4 \sqrt{J_\parallel D}~ \langle|S_{i}^z|\rangle \label{JDS}
\end{equation}

Here $J_\parallel$ is the intrachain exchange. From Eq.
(\ref{JDS}), using the measured value of $\omega_0/2\pi=80$~GHz
and the values of $D$ (see above) and $J_\parallel = $~9~meV
determined in Ref.~\onlinecite{Uchiyama}, we estimate
$\langle |S_{z}^i| \rangle \approx$~0.06 for $x=0.04$. The observed linear
concentration dependence of the AFMR gap is in accordance with
Eqs.~(\ref{averS}) and (\ref{JDS}).
 With the use of the measured value of $T_N$, and of relation~(\ref{theta}) we
estimate   $\xi \sim 10$, (here $J_\perp=0.11$~meV has been taken
from Ref.~\onlinecite{Uchiyama}). This evaluation of the
correlation length is in accordance with the value taken  above
for the analysis of the dependence of the paramagnetic resonance
linewidth on concentration  and with the correlation length of the
unperturbed Haldane chain which is known to be approximately 7
interspin distances.  \cite{Miyashita}

\section{Conclusion}
We identified the  defects created in the Haldane magnet
PbNi$_2$V$_2$O$_8$ via the diamagnetic dilution by means of ESR.
The absence of a zero-field splitting in the paramagnetic spectra
confirms the effective $S=1/2$ degrees of freedom created by
breaking $S=1$  spin chains.  Antiferromagnetic resonance modes
were found for different doping concentrations below the ordering
temperatures.  The dependence of the antiferromagnetic resonance
gap on the impurity concentration is similar to that of the N\'eel
temperature.  At low doping concentration the coexistence of
paramagnetic and antiferromagnetic resonance modes was found. It
is interpreted as a microscopic phase separation of the magnetic
phases at the N\'{e}el point analogous to the phase separation in
the doped spin-Peierls magnet.\cite{Glazkov3}   The correlation
length is estimated to be about 10 interspin distances.  At low
doping the N\'{e}el temperature and the antiferronagnetic
resonance gap were found to be proportional to the concentration
of the impurities in accordance with the molecular-field theory.

\begin{acknowledgments}
We kindly acknowledge D.~Vieweg for SQUID measurements. The work
was supported  by the Russian  Foundation for Basic Research
(RFBR), grant No.00-02-17317, by joined grant of RFBR and Deutsche
Forschungs Gemeinschaft (DFG),
 by the INTAS
 project No.  99-0155, by Award No.  RP1-2097 of
the U.S.  Civilian Research and Development Foundation for the
Independent States of the former Soviet union (CRDF), by
Bundesministerium f\"{u}r Bildung und Forschung under contract No.
13N6917(EKM), and DFG via Sonderforschungsbereich 484.

\end{acknowledgments}

\end{document}